\documentclass[aps,showpacs,nofootinbib,prd,twocolumn]{revtex4-1}
\usepackage{graphicx}
\usepackage{amssymb}
\usepackage{amsmath}
\usepackage[svgnames]{xcolor}
\usepackage{mathtools,slashed}
\usepackage{epstopdf}
\usepackage[utf8]{inputenc}
\usepackage{url}
\usepackage[colorlinks,citecolor=DarkGreen,linkcolor=DarkRed,urlcolor=DarkBlue]{hyperref}

\usepackage{newtxtext}
\usepackage{newtxmath}

\usepackage{epsfig}

\newcommand{\be}{\begin{eqnarray}}
\newcommand{\ee}{\end{eqnarray}}
\begin{document}

\title{Finite volume and magnetic field effects on the two-pion correlation function in relativistic heavy-ion collisions}


\author{Alejandro Ayala$^{1,2}$}
\author{Santiago Bernal-Langarica$^1$}
\author{Cristian Villavicencio$^3$}
  \address{
  $^1$Instituto de Ciencias
  Nucleares, Universidad Nacional Aut\'onoma de M\'exico, Apartado
  Postal 70-543, CdMx 04510,
  Mexico.\\
  $^2$Centre for Theoretical and Mathematical Physics, and Department of Physics,
  University of Cape Town, Rondebosch 7700, South Africa.\\
  $^3$Centro de Ciencias Exactas and Departamento de Ciencias B\'asicas, Facultad de Ciencias, Universidad del B\'io-B\'io, Casilla 447, Chill\'an, Chile.
  }
\begin{abstract}

We study the combined effects of a finite volume and an external magnetic field on the charged two-pion correlation function. For these purposes, we consider a dilute system of pions where the finite volume effects are introduced computing the pion wave functions with rigid boundary conditions in a cylindrical geometry in the presence of a uniform and constant magnetic field. We find that for slow pions, namely, for the case where the average pair momentum is small, the correlation function shows a large distortion, as opposed to the case where the average pair momentum is large. For a finite density system, the intercept of the correlation function is reduced, signaling the increasing importance of the pion ground state contribution. An increasing strength of the magnetic field reduces the importance of the ground state and the intercept becomes closer to 2.

\end{abstract}
\maketitle

\section{Introduction}

The properties of relativistic pion systems have attracted attention over several years in different contexts, ranging from the early universe~\cite{Ebert:2008tp,Vovchenko:2020crk} and astrophysics~\cite{Umeda:1994it,Takahashi:2007qu} to relativistic heavy-ion collisions~\cite{Voskresensky:1996ur,Florkowski:1995hv,Begun:2015ifa}. One of the main motivations has been the search for conditions that can lead to the production of a Bose-Einstein Condensate (BEC) and its possible signatures~\cite{Kogut:2001id,Loewe:2002tw,Greiner:1993jn,Zimanyi:1979ga,Loewe:2004mu,Loewe:2005yn,Andersen:2007qv,Ebert:2011tt,Mammarella:2015pxa,Zhang:2015baa,He:2005nk,He:2006tn,Sun:2007fc,Mu:2010zz,Ayala:2012dk}. It has been recently shown that if the system can be regarded as having a finite volume, the relative contribution from the condensate state to the system properties is enhanced~\cite{Ayala:1997ip,Begun:2008hq}. The fireball produced during the evolution of a heavy-ion reaction can be regarded as providing this finite volume. More recently, it has also been realized that a magnetic field of considerable strength, albeit short lived, can be produced in semi-central heavy-ion collisions~\cite{2008NuPhA.803..227K,MCLERRAN2014184,2009IJMPA..24.5925S}. A natural question that emerges is what is the effect of the magnetic field produced in this kind of collisions on the formation of the condensate in a finite volume system. This question has been partially addressed in Ref.~\cite{Ayala:2016awt} where we found that the critical temperature for the formation of the condensate, when a magnetic field is present, turns out to be larger than the one obtained by considering only finite
size effects. 

Experimental signatures of pion condensation are more easily accessible from measurements of two-pion correlation functions~\cite{ALICE:2015ryj}. The height of the intercept parameter can be linked to the fraction of pions that are found in the ground state. Recall that if a sizable fraction of pions is condensed, these can be regarded as being emitted from the fireball in a coherent state. Therefore, their contribution can be singled out from the contribution of the rest of the states. The presence of a condensate shows itself in terms of an intercept of the two-pion correlation function smaller than 2 at vanishing relative momentum. Finite size effects on the height of this intercept, when a sizable fraction of pions are in the lowest energy state, have been studied in Ref.~\cite{Ayala:2001pf}. In this work we include the effects of a magnetic field and study how this intercept varies, when the field strength, the temperature, the density and the system size are varied. We consider a non-equilibrium chemical potential to account for a given number of charged pions~\cite{Begun:2015ifa}. The work is organized as follows: In Sec.~\ref{secII} we formulate the way the finite volume pion states can be obtained in the presence of a magnetic field. This is accomplished by finding the solutions of the Klein-Gordon equation with rigid boundaries, minimally substituting the canonical by the kinematical momentum. We work within a cylindrical finite volume to allow for different transverse and longitudinal (with respect to the magnetic field) dimensions. Armed with the explicit eigenfunctions, in Sec.~\ref{secIII} we study the magnetic field and finite volume combined effects on the two-pion correlation function. Finally, we summarize and conclude in Sec.~\ref{concl}.

\section{Two-pion correlation function in a finite volume and within a magnetic field }\label{secII}

We study a dilute system of charged pions contained within a cylindrical finite volume of radius $R$ and height $L$, immersed within a constant magnetic field $\vec{B}$ in the $\hat{z}$ direction. We include the magnetic field effects by means of the minimal substitution $\vec{p} \to \vec{p} + q \vec{A}$, where $q$ is the pion charge. Working in the symmetric gauge,  $\vec{A} = \tfrac{|\vec{B}|}{2} (-y, x, 0)$, the system of pions can be described in terms of the Klein-Gordon equation
\begin{equation}
    \label{KGeq}
    \left\{ -\left( i \frac{\partial}{\partial t} \right)^2 + \left[-i \nabla + q \vec{A}\right]^2 + m^2\right\} \psi(\vec{r},t) = 0,
\end{equation}
where $m$ is the pion mass. To implement the finite volume effects, we look for the stationary states subject to the boundary conditions
\begin{eqnarray}
    \label{BCeq}
    \psi(r = R,t) & = & 0 , \\
    \psi(z=\pm \tfrac{L}{2},t) & = & 0.
\end{eqnarray}

The stationary states are given by
\begin{eqnarray}
    \label{WaveFuncteq}
    \psi_{n l j} (r,\theta,z,t) & = & \frac{A_{nlj}}{\sqrt{2E_{nlj}}} e^{-i E_{nlj} t} e^{-i l\theta}  \cos \left( k_j \,z\right)\nonumber\\ &\times&e^{-\frac{qBr^{2}}{4}} r^{l} {}_1 F_1 \left[-a_{nl} , l+1 ; \frac{qBr^2}{2}\right],
\end{eqnarray}
where the quantized momentum in $\hat z$ direction is defined as
\begin{equation}
    k_j\equiv  \frac{(2j +1)\pi}{L}
\end{equation}
with $j,l=0,1,2\ldots$, and where ${}_1 F_1$ is a confluent hypergeometric function.Notice that these states correspond to the well-known solution for a harmonic oscillator with rigid boundary conditions (see for example Ref.~\cite{Al-Hashimi:2012mnv}). The parameters $a_{nl}$ are obtained from the solutions of the boundary condition
\begin{eqnarray}
{}_1 F_1\left[-a_{nl} , l+1 ; \frac{qBR^2}{2}\right]=0,
\label{1F1roots}
\end{eqnarray}
and are related to the energy eigenvalues $E_{nlj}$ by
\begin{equation}
    a_{nl} = \frac{E_{nlj} ^2 - m^2 - k_j^2}{2qB} - \frac{2l+1}{2}.
\label{eigenvalues}
\end{equation}
Level quantization in the $\hat{z}$ direction corresponds to the quantization of a one-dimensional rigid box of length $L$ and is described by the quantum number $j$. Level quantization in the $\hat{x}-\hat{y}$ direction is achieved from the solutions of Eq.~(\ref{1F1roots}) and accounts both for the presence of the magnetic field as well as for the finite size of the cylinder of radius $R$.
The energy in the previous equation can be written then in terms of the parameter $a_{nl}$
\begin{equation}
    E_{nlj}^2=k_j^2+m^2+qB(2l+1+2a_{nl}),
\end{equation}
from where the Landau levels can be identified in terms of the quantum numbers $n$ and $l$. The quantities $A_{nlj}$ in Eq.~\eqref{WaveFuncteq} are the normalization constants and are obtained from the condition
\begin{equation}
    \label{Normalizationeq}
    \int d^{3}r \, \psi_{n l j} ^{*} (\vec{r},t) \overleftrightarrow{\frac{\partial}{\partial t}} \psi_{n l j} (\vec{r},t)=  1 .
\end{equation}
Equation~\eqref{WaveFuncteq}, together with the energy eigenvalues obtained from Eq.~\eqref{eigenvalues}, constitute the set of properly normalized eigenfunctions in terms of which the various multiparticle distributions can be expressed.

In order to describe the system near equilibrium, we consider a thermal statistical distribution in the gran canonical ensemble. Following Ref.~\cite{Ayala:2001pf}, the corresponding occupation number $N_\lambda$ for a given state is 
\begin{equation}
    N_\lambda = 
    \frac{1}{\exp (E_\lambda - \mu)/T - 1},
    \label{occupation}
\end{equation}
where 
$\lambda$ represents the set of quantum numbers $\{ n,\ l,\ j \}$, $T$ is the system temperature and $\mu$ the chemical potential associated to the pion number density. The situation we consider corresponds to a charge balanced system with equal numbers of positive and negative pions.
Since from the strong interaction perspective, there is no difference between positive and negative charge pions, both kinds of particles need to be considered as populating the quantum levels, given that in the absence of particle interactions other than with the external field (dilute system approximation), the magnetic effects enter with the absolute value of the particle's charge. Nevertheless, the correlation function refers only to indistinguishable particles, that is either positive or negative charged pions. Notice that since the chemical potential does not correspond to a strictly speaking conserved charge, it cannot be included in the Hamiltonian and thus in the Lagrangian in the same way that a conserved charge would be included. In this sense, our chemical potential corresponds to an effective description of the (in average) approximately conserved number of charged pions. 

Let  $\psi_\lambda (\vec{p})$ denote the Fourier transformed wave function for the state with quantum numbers $\lambda$, namely,
\begin{equation}
    \psi_\lambda (\vec{p}) = \int d^3 r\, e^{-i\vec{p}\cdot\vec{r}}\psi_\lambda (\vec{r}).
\end{equation}
Accounting for the normalization in Eq.~\eqref{WaveFuncteq}, the single-pion momentum distribution can be written as 
\begin{equation}
    \label{MomDistreq}
    P_1 (\vec{p}) \equiv \frac{d^3 N}{d^3 p} = \frac{1}{(2\pi)^3} \sum_{\lambda} 2 E_\lambda N_\lambda \psi_\lambda ^* (\vec{p}) \psi_\lambda (\vec{p}).
\end{equation}
The total number of pions can be shown to be obtained from
\begin{equation}
N = \sum_\lambda \frac{1}{\exp (E_\lambda - \mu)/T - 1}.
\label{TotalNumber}
\end{equation}

For a totally chaotic pion source, the two-pion distribution is given by
\begin{eqnarray}
    \label{2pionMomDistreq}
    \!\!\!\!\!\!\!\!P_2 (\vec{p}_1, \vec{p}_2)  &\equiv& \frac{d^6 N}{d^3 p_1 d^3 p_2} =  P_1 (\vec{p}_1) P_1 (\vec{p}_2)\nonumber\\ 
    &+& \left\vert\frac{1}{(2\pi)^3} \sum_{\lambda} 2 E_\lambda N_\lambda \psi_\lambda ^* (\vec{p}_1,t) \psi_\lambda (\vec{p}_2,t)\right\vert^2\!\!,
\end{eqnarray}
from where the two-pion correlation function $C_2$ can be expressed in terms of the one and two-pion momentum distributions as 
\begin{eqnarray}
    \!\!\!\!\!\!\!\!C_2 (\vec{p}_1, \vec{p}_2) &=& \frac{P_2 (\vec{p}_1,\vec{p}_2)}{P_1 (\vec{p}_1) P_1 (\vec{p}_2)}\nonumber\\ 
    &=&1 + \frac{\displaystyle{\left\vert \sum_\lambda  E_\lambda N_\lambda \psi_\lambda ^* (\vec{p}_1) \psi_\lambda (\vec{p}_2)\right\vert^2}}{\displaystyle{\sum_\lambda  E_\lambda N_\lambda \left\vert \psi_\lambda (\vec{p}_1)\right\vert^2 \sum_\lambda  E_\lambda N_\lambda \left\vert \psi_\lambda (\vec{p}_2)\right\vert^2}}.\nonumber\\
    \label{corrchaotic}
\end{eqnarray}

When the pion source is not totally chaotic, which happens when the fraction of pions in the ground state is not negligible, we must separate the contribution of the ground state to the correlation function~\cite{Ayala:2001pf} as
\begin{eqnarray}
    \!\!\!C_2 (\vec{p}_1, \vec{p}_2) &=&  1 + \frac{\displaystyle{\left\vert \sum_{\lambda\neq \lambda_0}  E_\lambda N_\lambda \psi_\lambda ^* (\vec{p}_1) \psi_\lambda (\vec{p}_2)\right\vert^2 }}{\displaystyle{\sum_\lambda  E_\lambda N_\lambda \left\vert \psi_\lambda (\vec{p}_1)\right\vert^2 \sum_\lambda  E_\lambda N_\lambda \left\vert \psi_\lambda (\vec{p}_2)\right\vert^2}}\nonumber\\
    &+&
    \frac{\displaystyle{\left\vert\frac{}{} E_{\lambda_0} N_{\lambda_0} \psi_{\lambda_0} ^* (\vec{p}_1) \psi_{\lambda_0} (\vec{p}_2)\,\right\vert^2}}{\displaystyle{\sum_\lambda  E_\lambda N_\lambda \left\vert \psi_\lambda (\vec{p}_1)\right\vert^2 \sum_\lambda  E_\lambda N_\lambda \left\vert \psi_\lambda (\vec{p}_2)\right\vert^2}},
    \label{modcorr}
\end{eqnarray}
where $\lambda_0$ represents the set of quantum numbers corresponding to the ground state. Notice that when pions coming from the ground state are treated separately from the ones coming from the excited states, the correlation function $C_2(\vec{p}_1, \vec{p}_2)$ fails to reach its maximum possible value $C_2(0)_{ max}=2$ as the ground state occupation increases~\cite{Ayala:2001pf}. This happens since for $\vec{q}=0$, the numerator and denominator in the second and third terms on the right-hand side of Eq.~(\ref{modcorr}) are no longer equal, as they were in Eq.~(\ref{corrchaotic}) for the case of a totally chaotic source.

The explicit expression for the Fourier transformed wave function is
\begin{eqnarray}
    \psi_\lambda (\vec{p} ) &=& (-\:i)^l e^{-\: i l \theta_i} \frac{(2\pi)^2 R^{l+2} A_\lambda}{\sqrt{2E_\lambda}} \frac{(-1)^j (2 j+1) L \cos \left(\frac{L p_{z}}{2}\right)}{(2 j + 1 )^2 \pi^2 - L^2 p_{z} ^2}\nonumber\\ &\times&\int_0 ^1 dx\, x^{l+1} e^{-\:\frac{qBR^2 x^2}{4}} J_l (p Rx)\nonumber\\
    &\times&{}_1 F_1 \left[-a_{nl} , l+1; \frac{qBR^2}{2} x^2\right] ,
    \label{explFT}
\end{eqnarray}
where in cylindrical coordinates $\vec{p} = (p,\theta,p_z)$.

With the expressions for the eigenfunctions in momentum space at hand, we can study the combined effects of the magnetic field and finite volume on the correlation function. We now proceed to show these properties. 

\section{Magnetic field and finite volume effects}\label{secIII}

The two-pion correlation function $C_2$ depends on the six kinematical variables corresponding to the two-pion momenta as well as parametrically on $\mu$, $T$, $|qB|$, $R$ and $L$. The properties of this correlation function can be more easily studied by setting particular configurations for these kinematical variables. In this work we consider the behavior of $C_2$ as a function of the magnitudes of the momentum difference,  in the transverse $q=|p_1-p_2|$ and longitudinal $q_z=|p_{z1}-p_{z2}|$ directions, and for fixed values of the average 
transverse $P=\tfrac{1}{2}\left(p_1+p_2\right)$ and longitudinal $P_z=\tfrac{1}{2} (p_{z1}+p_{z2})$ momenta, for different values of the external parameters. 

In order to make contact with the previous findings of Ref.~\cite{Ayala:2001pf}, we first consider the case when $|qB|=0$. This can be accomplished from Eq.~\eqref{WaveFuncteq} taking the limit $\vert qB \vert \to 0$ using the identity \cite{abramowitz+stegun}:
 \begin{equation}
 \lim_{\alpha \to \infty} \frac{{}_1 F_1 (\alpha, b, - \frac{z}{\alpha})}{\Gamma (b)} = z^{\frac{1}{2}(1 - b)} J_{b-1} (2\sqrt{z}),
 \end{equation}
whereby we obtain
 \begin{eqnarray}
     \psi_{nlj} (\vec{r},t) &=& \frac{\tilde{A}_{\lambda}}{\sqrt{2E_{nlj}}}e^{-\: i E_{nlj} t} e^{-\: il\theta} \cos \left(\frac{(2j+1)\pi}{L}z\right)\nonumber\\
     &\times&J_l \left( \sqrt{E_{nlj} ^2 -m^2 -\frac{(2j +1)^{2}\pi^{2}}{L^{2}}} \, r \right) ,
     \label{qB0WaveFunc}
 \end{eqnarray}
where $\Tilde{A}_\lambda$ is computed using the normalization condition in Eq.~\eqref{Normalizationeq}, $J_l$ is a Bessel function and the energy eigenvalues are obtained from the condition
\begin{equation}
     J_l \left(\sqrt{E_{nlj} ^2 - m^2 - \frac{(2j +1)^{2}\pi^{2}}{L^{2}}}\ R\right) = 0.
\end{equation}
Notice that Eq.~\eqref{qB0WaveFunc} can also be directly obtained from Eq.~\eqref{KGeq}, by setting the vector potential $\vec{A} = 0$, namely, by directly solving the Klein-Gordon equation in the absence of an external field, albeit still obeying cylindrical boundary conditions.

\subsection{Zero chemical potential}

\begin{figure}[t]
    \centering
    \includegraphics[width=0.475\textwidth]{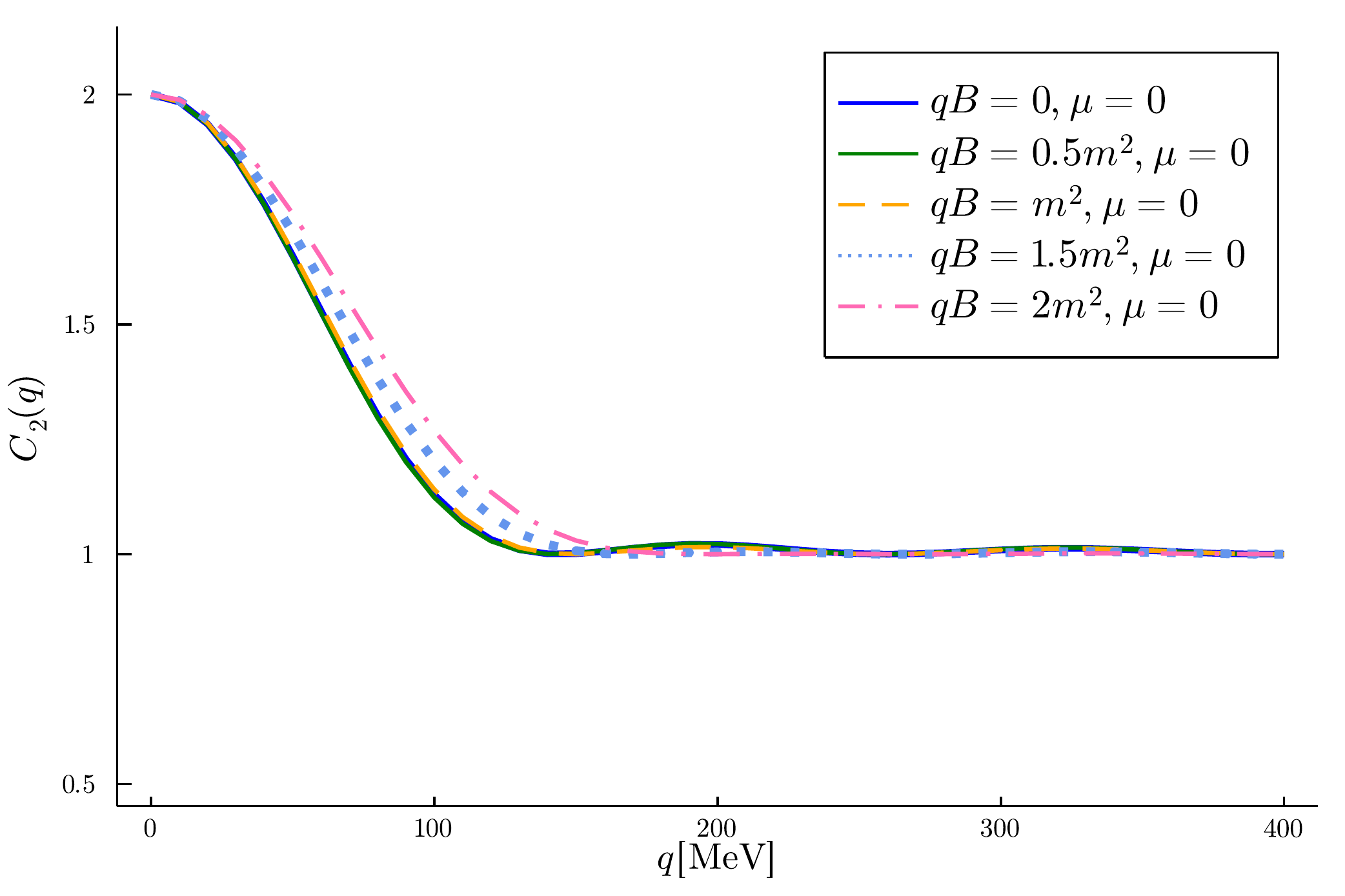}
     \caption{$C_2 (q)$ for fixed values of $P = 500$ MeV, $p_{z1} = p_{z2} = 0$ and $\theta_2 = \theta_1$, $R = 5$ fm and $L = 10$ fm. 
     In all cases the temperature and chemical potential have been held fixed to $T = 100$ MeV and $\mu = 0$. Notice that the increment of the correlation function width is appreciable only for the highest considered field strength.}
     \label{FigC2_mu0_difqB}
\end{figure}

In order to have a benchmark reference to compare against the case describing a finite density pion system, we hereby first consider the case where the number of pions is not fixed and thus of a vanishing chemical potential.
Figure~\ref{FigC2_mu0_difqB}, shows the behavior of $C_2 (q)$ for fixed values of $P = 500$ MeV, $p_{z1} = p_{z2} = 0$ and $\theta_2 = \theta_1$, taking $R = 5$ fm and $L = 10$ fm and for different magnetic field strengths, $qB = 0,\; 0.5\,m^2,\; m^2,\; 1.5\,m^2 \text{ and } 2 \, m^2$. The temperature and chemical potential have been held fixed to $T = 100$ MeV and $\mu = 0$, respectively. Notice that changes on the width of the correlation function start appearing only when $\vert qB \vert \gtrsim m^2$. 

\begin{figure}[b]
    \centering
    \includegraphics[width=0.475\textwidth]{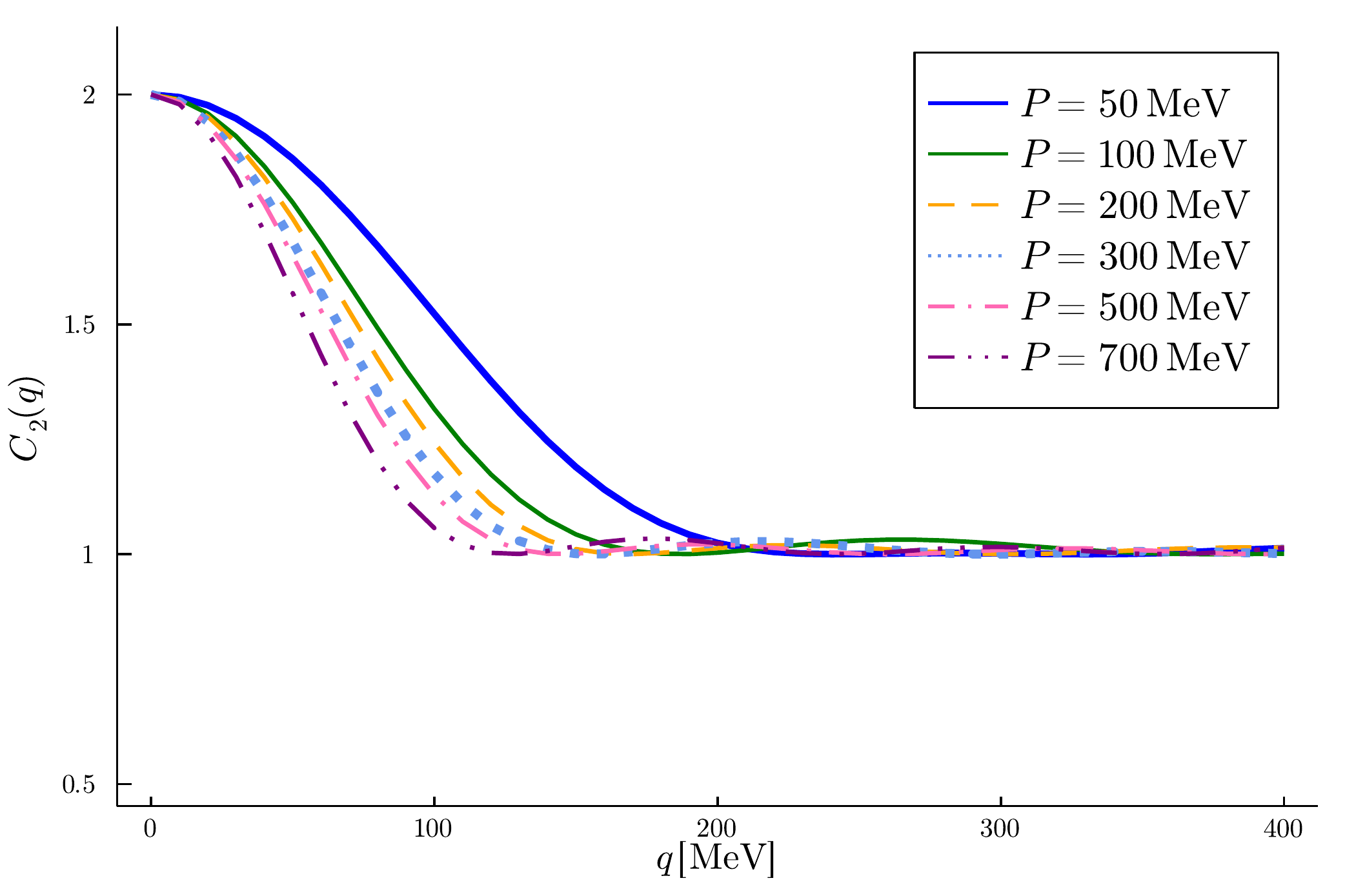}
     \caption{$C_2 (q)$ for fixed values of $\vert qB\vert = 0$, $p_{z1} = p_{z2} = 0$ and $\theta_2 = \theta_1$, $R = 5$ fm and $L = 10$ fm. 
     In all cases the temperature and chemical potential have been held fixed to $T = 100$ MeV and $\mu = 0$. Notice that the width of the correlation function is a monotonically decreasing function of $P$.}
     \label{FigC2_qB0_difP}
\end{figure}

\begin{figure}[t]
    \centering
    \includegraphics[width=0.475\textwidth]{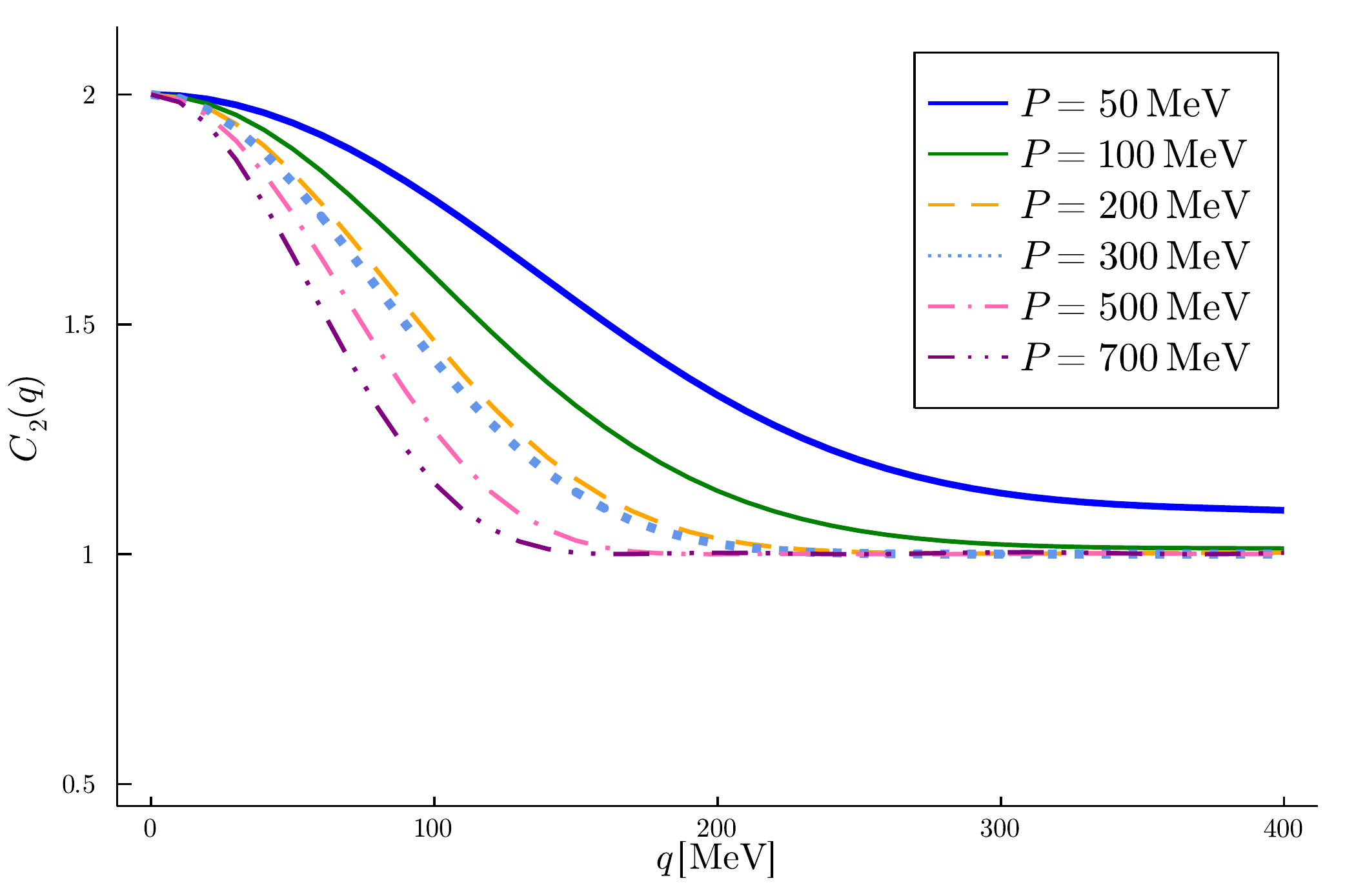}
     \caption{$C_2 (q)$ for fixed values of $\vert qB\vert = 2 \, m^2$, $p_{z1} = p_{z2} = 0$ and $\theta_2 = \theta_1$, $R = 5$ fm and $L = 10$ fm. 
     In all cases the temperature and chemical potential have been held fixed to $T = 100$ MeV and $\mu = 0$. Notice that the width of the correlation function is a monotonically decreasing function of $P$, however, for each of the considered values of $P$, the width is larger than for the corresponding case for $|qB|=0$.}
     \label{FigC2_qB2_difP}
\end{figure}

Figures~\ref{FigC2_qB0_difP} and~\ref{FigC2_qB2_difP} show the behavior of $C_2 (q)$ for different average momenta  and for the same set of parameters as in Fig.~\ref{FigC2_mu0_difqB} for a magnetic field strength $\vert qB \vert = 0$ and $\vert qB \vert = 2\,m^2$, respectively. Notice that in both cases the width of the correlation function is a monotonically decreasing function of $P$. A finite value of the field strength increases the width of the correlation function but does not change its monotonically decreasing trend with $P$.

\begin{figure}[b]
    \centering
    \includegraphics[width=0.475\textwidth]{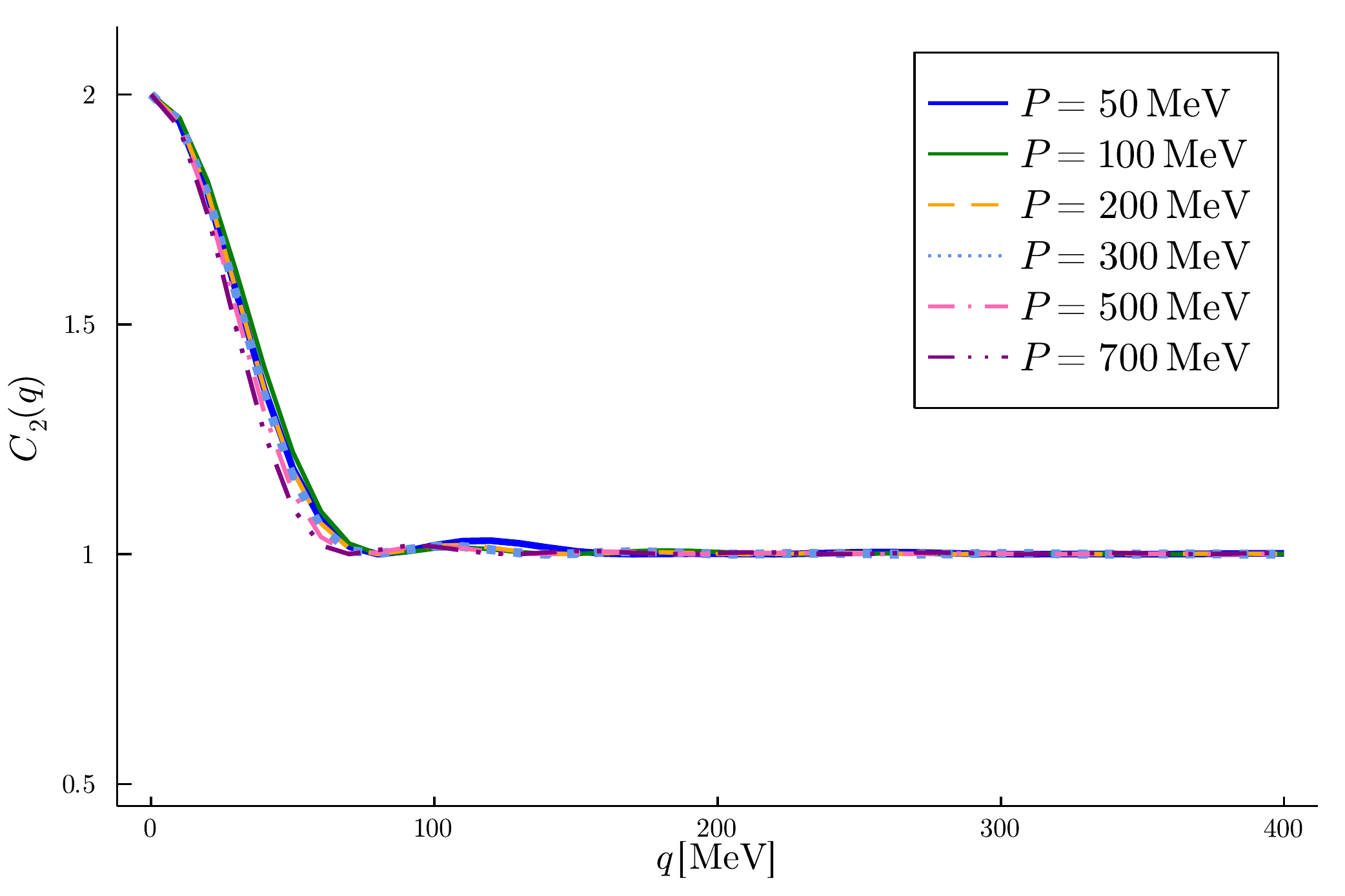}
    \caption{$C_2 (q)$ for fixed values of $\vert qB\vert = 0$, $p_{z1} = p_{z2} = 0$ and $\theta_2 = \theta_1$, $R = 10$ fm and $L = 5$ fm.
    In all cases the temperature and chemical potential have been held fixed to $T = 100$ MeV and $\mu = 0$. Notice that the width of the correlation function is not a monotonic function of $P$.
    }
    \label{FigC2_qB0_difP_R10_L5}
\end{figure}
\begin{figure}[t]
    \centering
    \includegraphics[width=0.475\textwidth]{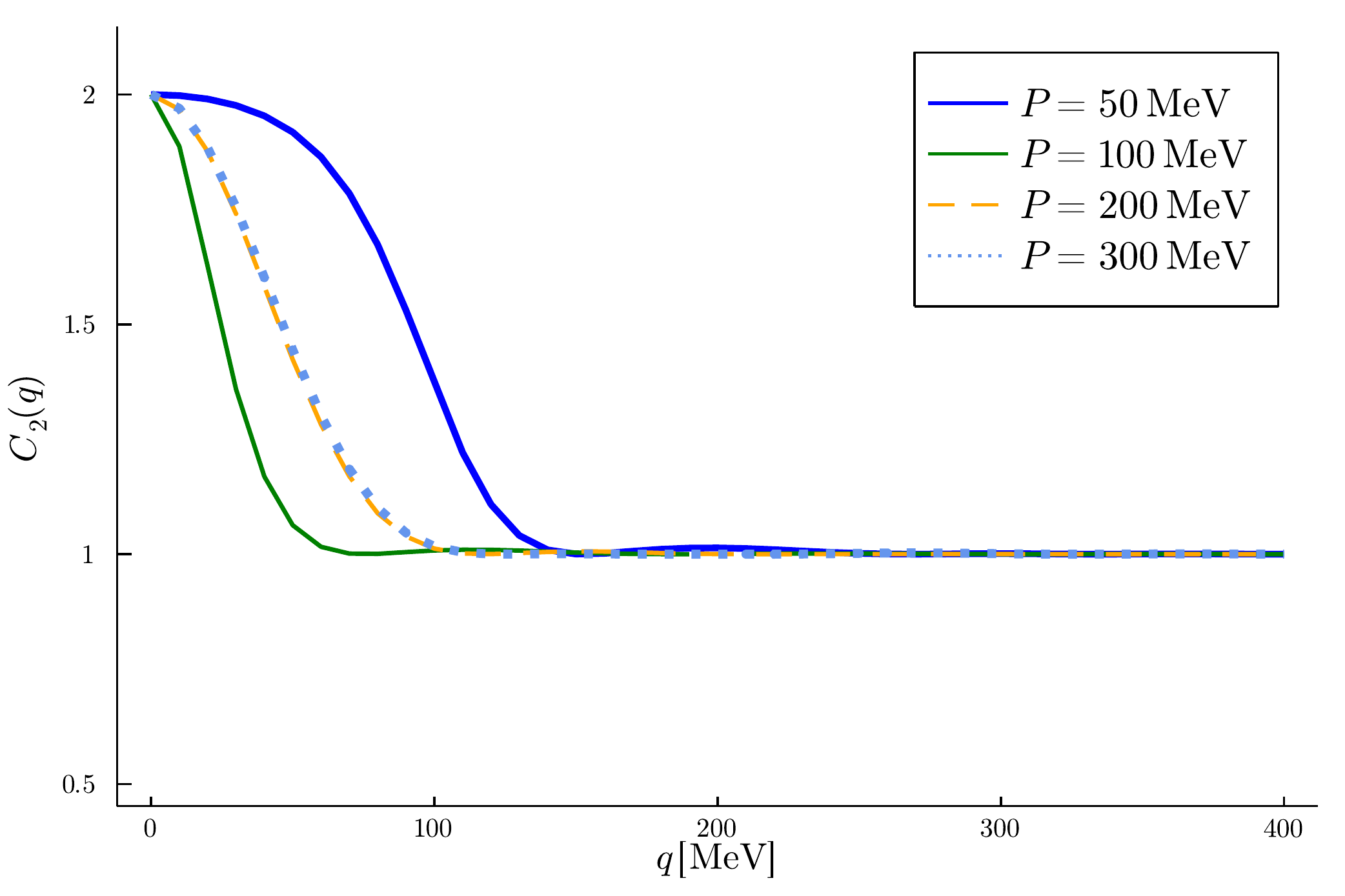}
    \caption{$C_2 (q)$ for fixed values of $\vert qB\vert = 2 m^2$, $p_{z1} = p_{z2} = 0$ and $\theta_2 = \theta_1$, $R = 10$ fm and $L = 5$ fm. 
    In all cases the temperature and chemical potential have been held fixed to $T = 100$ MeV and $\mu = 0$. Notice that the width of the correlation function is not a monotonic function of $P$.
    }
    \label{FigC2_qB2_difP_R10_L5}
\end{figure}

Figures~\ref{FigC2_qB0_difP_R10_L5} and~\ref{FigC2_qB2_difP_R10_L5} show the behavior of $C_2(q)$ again for different average momenta, $p_{z1} = p_{z2} = 0$, $\theta_2 = \theta_1$, $T=100$ MeV and $\mu=0$, for magnetic field strengths $\vert qB \vert = 0$ and $\vert qB \vert = 2\ m^2$, respectively, but this time for $R=10$\,fm and $L=5$\,fm, namely, the inverse hierarchy as for the case of Figs.~\ref{FigC2_qB0_difP} and~\ref{FigC2_qB2_difP}. 
Notice that when $|qB|=0$, the width of the correlation function barely changes, although this time it is not anymore a monotonic function of $P$. 
The width starts growing to then decrease as a function of $P$. The situation changes for the case when $|qB|=2\ m^2$ where the magnetic field produces the width to first decrease to then increase and decrease again as $P$ increases. We thus see that when the cylindrical box length along its symmetry axis (which is the same as the direction of the field) is shorter than the radius, the magnetic field has a more dramatic effect on the width. For pions with a smaller $P$ the magnetic field produces a larger distortion of the correlation function. This behaviour signals that slower pions experience the field effects for a longer time, inducing a change on their individual directions of motion and thus affecting more significantly the correlation as a function of $q$. It is hereby pertinent to mention that in a peripheral relativistic heavy-ion collision, the case where $L>R$ is a closer description of the ellipsoid that corresponds to the geometry of interaction region than the case $R>L$. This happens because on average, the magnetic field is directed along the normal to the reaction plane and thus along the semi-major axis of this ellipsoid. 
We thus continue exploring the properties of the correlation function for $\mu\neq 0$ for the case $L>R$.

\subsection{Fixed pion number}

\begin{figure}[b]
    \centering
    \includegraphics[width=0.475\textwidth]{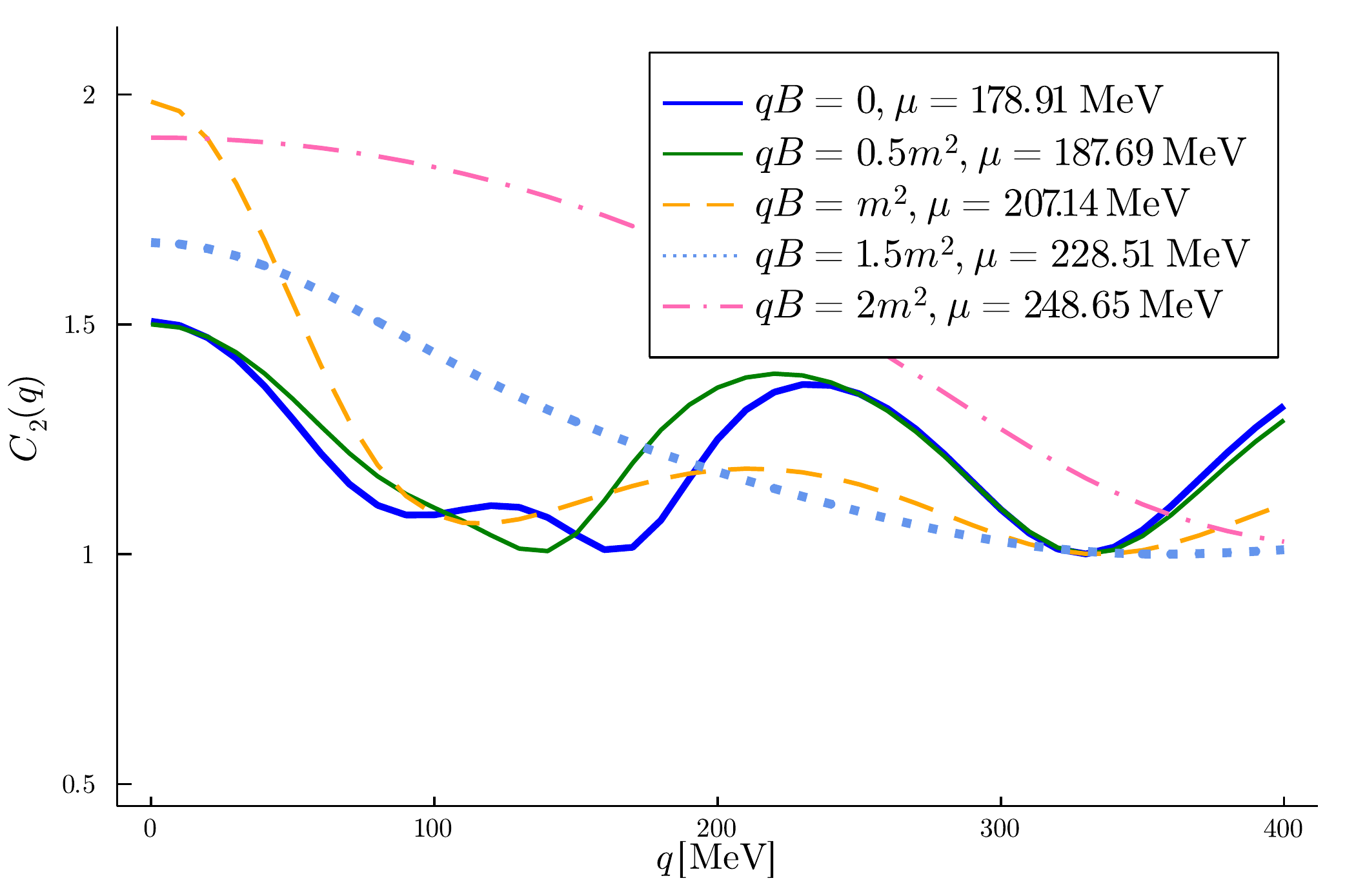}
    \caption{$C_2 (q)$ for fixed values of pion multiplicity $N = 320$, $p_{z_1} = p_{z_2} = 0$, $\theta_2 = \theta_1 = 0$, $R = 5$ fm and $L = 10$ fm. 
    In all cases the temperature and  the pair average momenta have been held fixed to $T = 100$\,MeV and $P = 300$\,MeV, respectively. 
    }
    \label{FigC2_N320_P300_T100}
\end{figure}

\begin{figure}[t]
    \centering
    \includegraphics[width=0.475\textwidth]{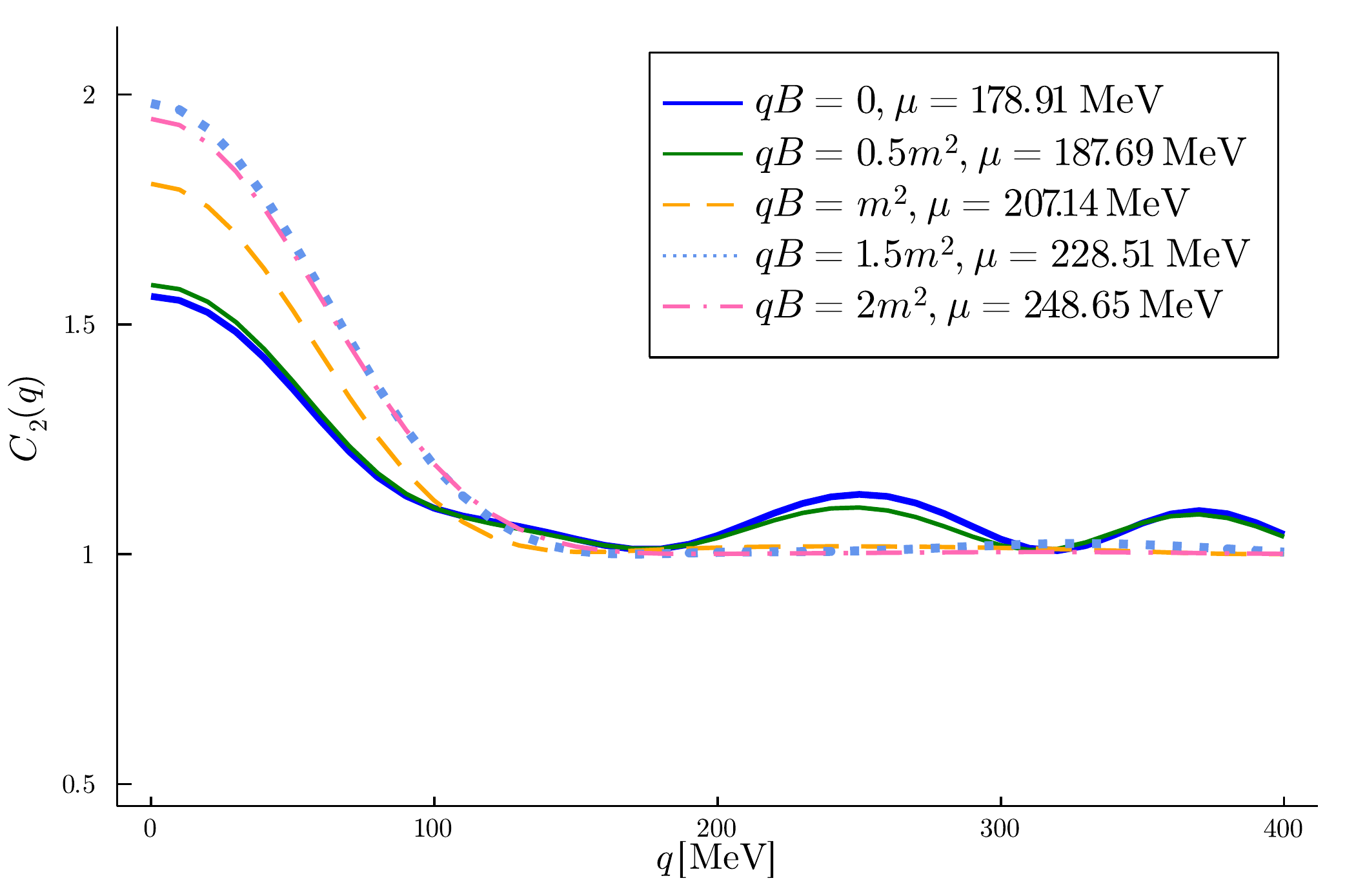}
    \caption{$C_2 (q)$ for fixed values of pion multiplicity $N = 320$, $p_{z_1} = p_{z_2} = 0$, $\theta_2 = \theta_1 = 0$, $R = 5$ fm and $L = 10$ fm. 
    In all cases the temperature and  the pair average momenta have been held fixed to $T = 100$\,MeV and $P = 500$\,MeV, respectively. 
    }
    \label{FigC2_N320_P500_T100}
\end{figure}

\begin{figure}[b]
    \centering
    \includegraphics[width=0.475\textwidth]{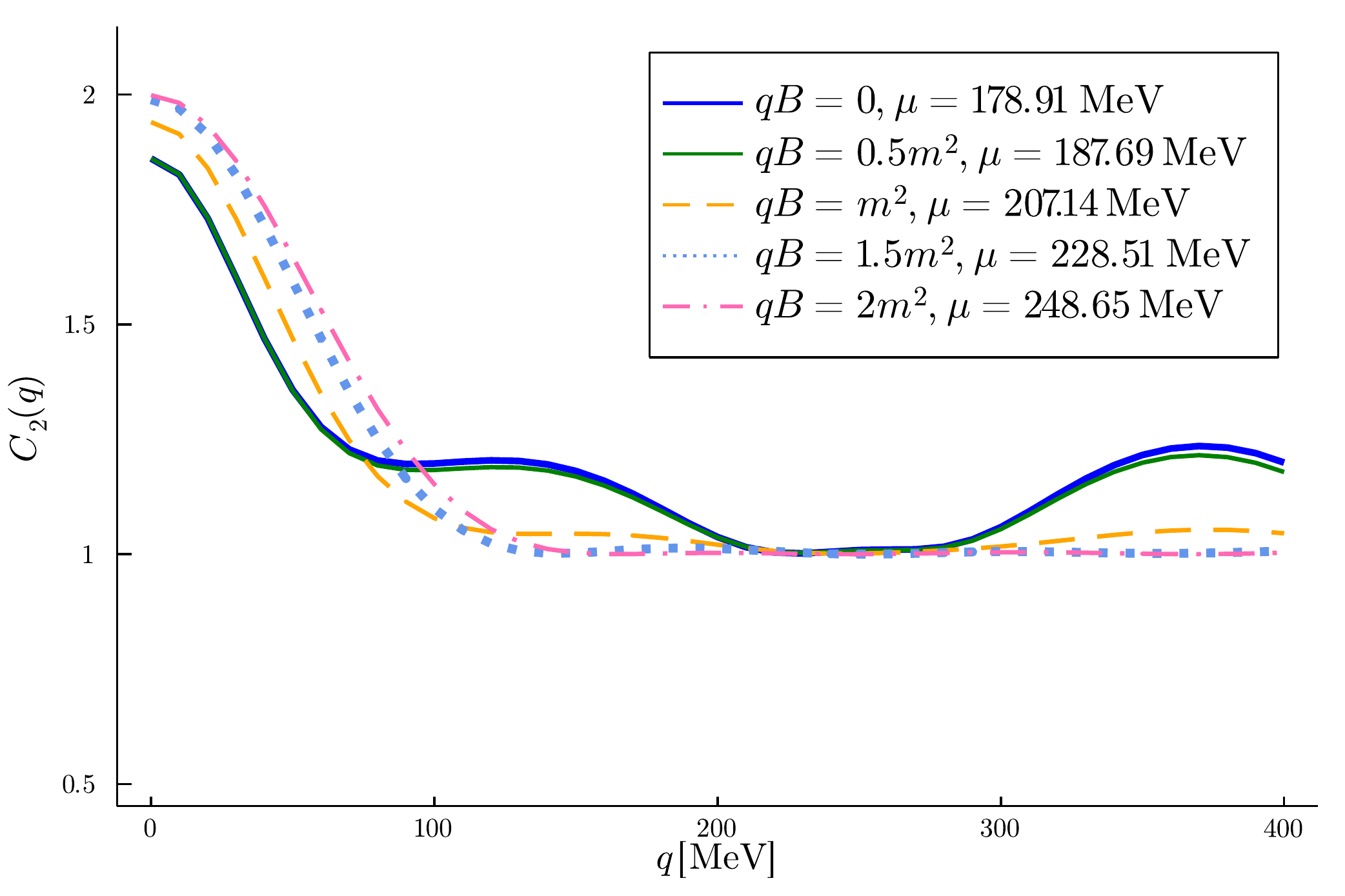}
    \caption{$C_2 (q)$ for fixed values of pion multiplicity $N = 320$, $p_{z_1} = p_{z_2} = 0$, $\theta_2 = \theta_1 = 0$, $R = 5$ fm and $L = 10$ fm. 
    In all cases the temperature and  the pair average momenta have been held fixed to $T = 100$\,MeV and $P = 700$\,MeV, respectively. 
    }
    \label{FigC2_N320_P700_T100}
\end{figure}

Figures \ref{FigC2_N320_P300_T100}, \ref{FigC2_N320_P500_T100} and \ref{FigC2_N320_P700_T100} show the behavior of $C_2 (q)$ for different magnetic field strengths $\vert qB\vert = 0,\; 0.5\,m^2,\; m^2,\; 1.5\,m^2,\; 2\,m^2$, $p_{z1} = p_{z2} = 0$, $\theta_2 = \theta_1$, $T=100$ MeV and $N = 320$, for different average pair momenta $P$ = 300, 500, 700 MeV. The chemical potential has been calculated for each case by fixing $N = 320$~\cite{Ayala:2016awt} from Eq.~\eqref{TotalNumber}. Notice that the energy eigenvalues $E_\lambda$ appearing in Eq.~(\ref{occupation}) are computed from the solutions of Eq.~(\ref{1F1roots}). When the particle number is fixed, the chemical potential adjusts itself to take on values such that, for the given temperature and system size, each energy level contributes with a finite occupation number such that the sum of each level contribution makes up for the total number of particles. When the system is dense, the difference between the ground state energy $E_0$ and the chemical potential becomes small but it never vanishes. This precludes the divergence of the ground state occupation number.

\begin{figure}[]
    \centering
    \includegraphics[width=0.475\textwidth]{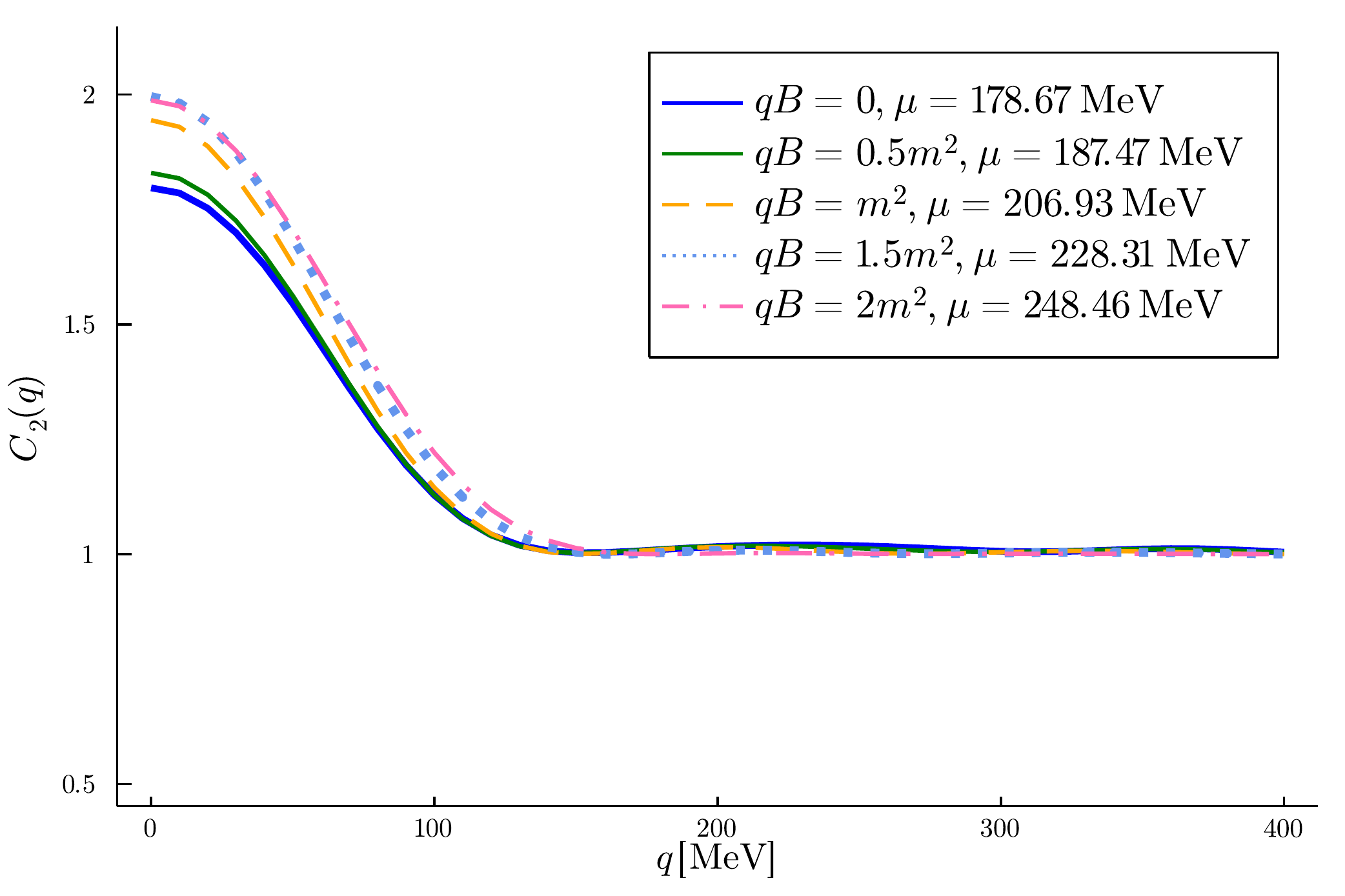}
    \caption{$C_2 (q)$ for fixed values of pion multiplicity $N = 320$, $p_{z_1} = p_{z_2} = 0$, $\theta_2 = \theta_1 = 0$, $R = 5$ fm and $L = 10$ fm. 
    In all cases the temperature and  the pair average momenta have been held fixed to $T = 150$\,MeV and $P = 500$\,MeV, respectively. 
    }
    \label{FigC2_N320_P500_T150}
\end{figure}
\begin{figure}[]
    \centering
    \includegraphics[width=0.475\textwidth]{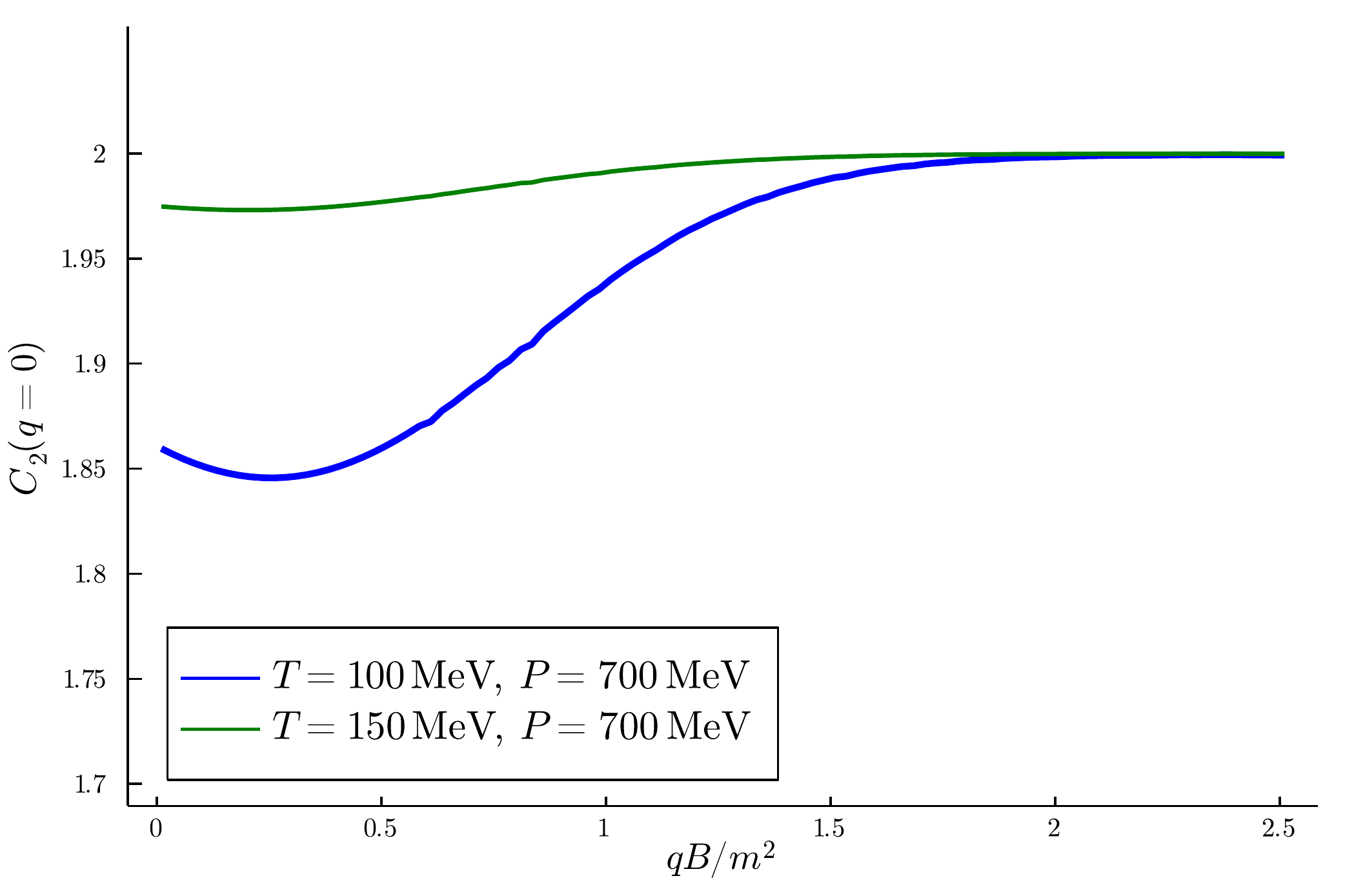}
    \caption{Correlation function intercept $C_2 (q=0)$ as a function of the field intensity for fixed values of the pion multiplicity $N = 320$, $p_{z_1} = p_{z_2} = 0$, $\theta_2 = \theta_1 = 0$, $R = 5$ fm and $L = 10$ fm. Notice that in both cases the intercept is not a monotonic function of the field intensity and tends to 2 for the largest field strengths shown.}
    \label{newfig}
\end{figure}

As before, pions with smaller $P$ show a correlation function with a more significant distortion. 
For small values of the magnetic field, this distortion shows up as an oscillation pattern that extends to larger values of $q$. 
These oscillations fade away for larger values of $|qB|$, however the distortion of the correlation function shows as an increase of the width and the intercept value. For finite values of $|qB|$, the latter effects are more significant for low momentum than for higher values or $P$. In classical terms, this behavior can be understood as coming from the relation between the cyclotron radius $R_0$ for a charged particle with momentum $p$ subject to the effect of a magnetic field $|qB|$, given by $R_0=p/|qB|$.
Although in the configurations we have studied, $p$ starts off in the radial direction, for low momentum pions the original direction of motion is bound to be changed to start orbiting the field lines. When $R_0<R$, that is, for slow pions, this change of direction produces larger effects. These properties can be summarized as follows: for a relatively low temperature and small values of $|qB|$ the population of the ground state is significant whereas for the same temperature, increasing values of $|qB|$ contribute to decrease the importance of this state. For finite values of the external field, low momentum pions are significantly more affected than fast moving ones. Notice that $C_2 (q = 0)$ is a non-monotonic function of $\vert qB \vert$. For the case of pions with larger $P$, this non-monotonic behavior is less pronounced but still persists. This effect is in agreement with the findings in Ref.~\cite{Lednicky:1999xz} where it is shown that when external effects are included, distortions of the correlation function are always more significant for slower pions. In all the cases discussed $C_2(0) < 2$, although for large values of $\vert qB \vert$, $C_2 (q = 0)\to 2$. In account of Eq.~(\ref{modcorr}), this shows that the condensate forms but tends to be destroyed for large magnetic field intensities.

Figure~\ref{FigC2_N320_P500_T150} corresponds to the case shown in Fig.~\ref{FigC2_N320_P500_T100} but considering a higher temperature $T=150$ MeV. As in the previous cases, the chemical potential has been calculated for each value of the magnetic field strength by fixing $N = 320$ from Eq.~\eqref{TotalNumber}. Notice that the non-monotonic behavior of $C_2 (q = 0)$ as a function of $\vert qB \vert$ is less evident but still persists. This happens because for higher temperatures, the contribution from the condensate becomes correspondingly less prominent. This is illustrated in Fig.~\ref{newfig} where we show $C_2(q=0)$ as a function of the field strength for two temperatures $T=100, 150$ MeV and $P=700$ MeV.

Overall, for the same number of pions in a given volume, for increasing strengths, the magnetic field produces that the intercept of the correlation function becomes closer to 2. We thus find that a large magnetic field produces a reduction of the contribution to the correlation function from the fraction of pions coming from the ground state. These findings are in agreement with the results of Ref.~\cite{Ayala:2016awt}, whereby the presence of a magnetic field increases the critical temperature for the formation of the condensate in a finite volume.
In all the studied cases, the destruction of the condensate is not sudden neither is monotonic as a function of the field intensity.

\section{Summary and conclusions}\label{concl}
In summary, we have studied the combined effects of a finite volume and magnetic field on the charged, two-pion correlation function. To include these effects, we have considered the dilute limit of the pion system computing their wave functions in the presence of the magnetic field assuming cylindrical symmetry with rigid boundary conditions.
To include the effects of a finite density, we introduce a chemical potential associated with the approximately (in average) conserved charged pion number. For a vanishing magnetic field and when the temperature is sufficiently small and the pion number is fixed (and thus the system is described with a non-vanishing chemical potential), $C_2(0) < 2$. The field greatly distorts the correlation function for pions whose average pair momentum is small. We interpret this behavior as coming from the deflection that pions experience due to the magnetic field, whose effect is larger for small pion momentum and more intense field strengths. Increasing values of the magnetic field reduce the contribution of the ground state producing that, for a finite density pion system, an intercept of the correlation function is closer to 2. 

Recall that for finite volume systems, the condensate forms for relatively large values of the chemical potential and for low temperatures and small volumes. Unlike the infinite volume case, where condensation occurs when the chemical potential reaches the pion mass (see Ref.~\cite{Begun:2015ifa}), in the finite volume case, there is not a precise critical temperature. As we have shown, the condensate formation is enhanced in the absence of a magnetic field, due to finite size effects. The increasing importance of the pion ground state contribution is signaled by the fact that $C_2(0) < 2$. This behavior has already been found in Ref.~\cite{Ayala:2001pf}. Although not monotonically, the  magnetic field tends to destroy the condensate.

Our findings emphasize that a magnetic field, such as the one produced in the early stages of a relativistic heavy-ion collision, when properly considered, can distort the correlation function, particularly for low momentum charged pions. This distortion may be missed in analysis that measure the correlation function for pions within a given solid angle that may not correspond to the ones carrying the original quantum correlations, due to the bending of the trajectory caused by the magnetic field. This effect can exclude from this solid angle some of the pions originally coming from regions close in phase space. Consequences of this effect as well as of the possible back-reaction currents produced by the charged pions in response to the external field, including a possible charged imbalanced pion system, are currently being pursued and will be reported elsewhere.

\begin{acknowledgements}
The authors thank A. Kisiel for useful suggestions. Support for this work was received in part by UNAM-DGAPA-PAPIIT grant number IG100219 and by Consejo Nacional de Ciencia y Tecnolog\'ia grant numbers A1-S-7655 and A1-S-16215. SBL was partially supported by an SNI-CONACYT fellowship.
C.V. acknowledges financial support from FONDECYT under grants 1190192 and 1200483.
\end{acknowledgements}

\bibliography{biblio}

\end{document}